\documentclass[fp]{jpsj3}
\usepackage{txfonts,color}

\title{Enhanced Diffusion of Molecular Motors in the Presence of 
Adenosine Triphosphate and External Force}

\author{Ryota Shinagawa and Kazuo Sasaki\thanks{E-mail: sasaki@camp.apph.tohoku.ac.jp}}
\inst{Department of Applied Physics, Graduate School of Engineering, Tohoku University, 
Sendai 980-8579, Japan
} %\\

\abst{
The diffusion of a molecular motor in the presence of a constant external force is 
considered on the basis of a simple theoretical model. 
The motor is represented by a Brownian particle moving in a series of 
parabolic potentials placed periodically on a line, 
and the potential is switched stochastically from one parabola to another 
by a chemical reaction, which corresponds to the hydrolysis or synthesis 
of adenosine triphosphate (ATP) in motor proteins.  
It is found that the diffusion coefficient as a function of the force exhibits 
peaks.    
The mechanism of this diffusion enhancement and the possibility of 
observing it in $\rm F_1$-ATPase, a biological rotary motor, are discussed. 
}

\begin{document}
\maketitle

%%%
\section{Introduction}
\label{sec:introduction}
Diffusion is a fundamental phenomenon in nonequilibrium physics 
and has been of research interest since Einstein~\cite{einstein1905, einstein 1956} 
figured out that the Brownian motion of a mesoscopic particle 
is caused by the fluctuation in the bombardment of fluid molecules on it. 
One area of recent interest is the \textit{diffusion enhancement} 
of a Brownian particle moving in a periodic potential. 
The diffusion of such a particle should be reduced  
compared with free diffusion (diffusion without a potential) 
because the potential barrier suppresses the meandering of the particle. 
The enhancement is possible when the system is driven away from 
the thermal equilibrium state, for example, 
by applying an external force~\cite{costantini99, reimann01, reimann02, speer12} 
or by switching the potential (an on-off ratchet mechanism)~\cite{zhou04, germs13}. 
The diffusion coefficient can be increased considerably  
by tuning the strength of a constant external 
force~\cite{costantini99, reimann01, reimann02}, 
the frequency of an unbiased ac force~\cite{ speer12}, 
or the rate of potential switching~\cite{zhou04, germs13}. 

In the case of a particle in a one-dimensional periodic potential 
in the presence of a constant external force $F$ 
(a particle in a \textit{tilted periodic potential}), 
the diffusion coefficient as a function of $F$ exhibits a peak at a value 
close to the maximum slope $F_\mathrm{c}$ of the potential~\cite{costantini99, 
reimann01, reimann02};  
%note that in the stationary state of the particle dynamics without thermal noise, 
note that in the absence of thermal noise, 
the particle remains stationary at a force-balanced location for $F < F_\mathrm{c}$ 
while it continues to run in one direction (``running state'') for $F > F_\mathrm{c}$. 
The reason for this diffusion enhancement is that the behavior of the particle,  
whether to remain in a potential well or to move to the adjacent well, 
is quite sensitive to thermal noise for $F$ close to $F_\mathrm{c}$, 
which results in a large dispersion of the particle displacement 
and hence a large diffusion coefficient. 
This diffusion enhancement was experimentally confirmed for a colloidal particle 
in a potential created by periodically arranged optical traps~\cite{lee06, blickle07, 
evstigneev08}. 
More recently, enhanced diffusion of this type has been observed for 
a rotary motor protein, $\rm F_1$-ATPase, 
in the \textit{absence} of ATP (a fuel molecule required for the motor to work) and 
in the presence of a constant external torque~\cite{hayashi15};  
from the dependence of the rotational diffusion coefficient on the torque,  
the potential barrier for the rotor of this motor was estimated for the first time. 

\begin{figure}[hbt]
\centerline{\includegraphics[width = 7.5cm]{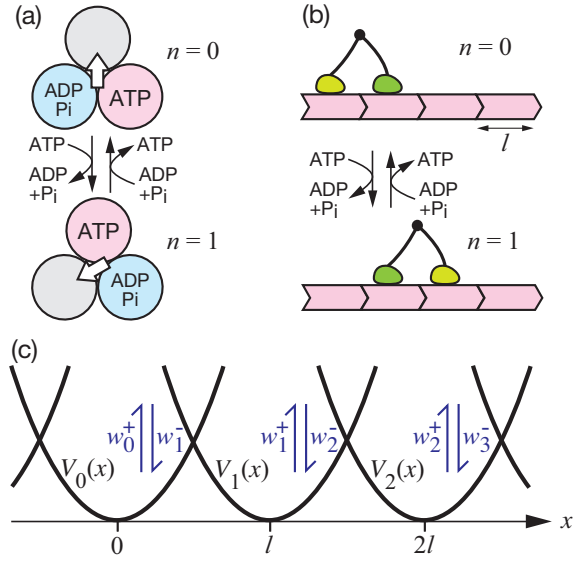}}
\caption{(Color online) 
	Model for motor protein that undergoes stepwise motion.
	A rotary motor (a) or a ``two-legged'' linear motor (b) is modeled as 
	a Brownian particle moving in a series of potential valleys (c) 
	arranged periodically (period $l$) on the $x$-axis. 
	Here, $x$ represents the rotation angle of the rotor (thick arrow) in (a) 
	or the displacement of the joint (big dot) connecting the legs in (b), 
	and $l$ corresponds to the step size. 
	See the text for a detailed explanation. 
}
\label{fig:model}
\end{figure}

In this work, we investigate the diffusion of molecular motors 
in the \textit{presence} of ATP and a constant external force on the basis of 
a simple theoretical model schematically shown in Fig.~\ref{fig:model} 
(see Sect.~\ref{sec:model} for a detailed description). 
It is found that the diffusion coefficient shows peaks in its dependence on the force; 
the mechanism of the diffusion enhancement at a high ATP concentration is 
similar to that for a particle in a tilted periodic potential, 
while an alternative mechanism applies when the ATP concentration is low. 
We also analyze the data on the torque dependence of the rotation rate 
for $\rm F_1$-ATPase~\cite{toyabe11} using our model, 
and predict that the diffusion enhancement of both types 
can be observed experimentally. 

The paper is organized as follows. 
In Sect.~\ref{sec:model}, our model for the molecular motor is defined, 
and the methods of calculating the average velocity and the diffusion coefficient of the motor 
are explained in Sect.~\ref{sec:vd}. 
The result for the dependence of the diffusion coefficient on the external force is 
presented and the mechanism of the diffusion enhancement is discussed 
in Sect.~\ref{sec:diffusion}. 
Section~\ref{sec:f1} is devoted to the analysis of $\rm F_1$-ATPase based on  
our model, and Sect.~\ref{sec:conclusion} provides concluding remarks. 

%%%
\section{Model}
\label{sec:model}
We consider the following model of a molecular motor, 
which can be a rotary motor 
consisting of a rotor and a stator [Fig.~\ref{fig:model}(a)] 
or a `two-legged' motor walking along a linear track [Fig.~\ref{fig:model}(b)]. 
Our model is analogous to the ones introduced in 
Refs.~\citen{zimmermann12, watanabe13, kawaguchi14}. 
Let $x$ be the rotation angle of the rotor for a rotary motor 
or the displacement of the joint of the two legs for a linear motor; 
hereafter, $x$ will simply be referred to as the displacement of the motor.  
We assume that the stator has a structure of $N$-fold symmetry for the rotary motor 
($N = 3$ for F$_1$-ATPase)  
or the track has a periodic structure for the linear motor. 
The motor changes its \textit{state}, denoted by an integer index $n = 
0, \pm1, \pm 2, \dots$, 
upon a chemical reaction catalyzed by the motor; 
$n$ is changed to $n + 1$ by a \textit{forward} reaction and to $n - 1$ by 
a \textit{backward} (reverse of the forward) reaction.  
Let $V_n(x)$ be the mechanical energy of the motor in state $n$, 
which represents the interaction between the rotor and the stator or 
the elastic energy of the legs together with the interaction 
between the motor and the track. 
The potentials $V_n(x)$ are assumed to satisfy the ``periodicity condition''
\begin{equation}
\label{eq:pcond}
	V_n(x) = V_0(x - n l)\,,
\end{equation}
where $l$ is $2\pi/N$ for the rotary motor or 
the period of the track for the linear motor [Fig.~\ref{fig:model}(c)]. 
It is assumed that $V_0(x) \to \infty$ as $|x| \to \infty$.  

The rates of transitions, $w_n^{\pm}(x)$, from state $n$ to $n \pm 1$ 
are assumed to depend on $x$ and possess the same periodicity as 
Eq.~(\ref{eq:pcond}), i.e., $w_n^\pm(x) = w_0^\pm(x - n l)$. 
Let $\Delta\mu$ be the free energy released by the forward reaction 
(the free energy of the environment is decreased by $\Delta\mu$ upon 
the reaction). 
Then, $w_n^{+}$ and $w_{n+1}^{-}$ should 
be related to each other through the detailed balance condition 
\begin{equation}
\label{eq:dbalance}
	\frac{w_{n+1}^{-}(x)}{w_n^{+}(x)}
	= \exp\left[\frac{V_{n+1}(x) - V_n(x) - \Delta\mu}{k_\mathrm{B}T}\right]\,, 
\end{equation}
where $k_\mathrm{B}$ is the Boltzmann constant and $T$ is the temperature 
of the environment. 

We assume that a constant external force $F$ is applied to the motor in addition to 
the force due to the interaction potential $V_n$. 
The fluid surrounding the motor also exerts forces on the motor: 
the drag force $-\gamma dx/dt$,  
with $\gamma$ being the frictional coefficient, and 
the random force, which is modeled as the Gaussian white noise. 
Then, the Fokker--Planck equation 
for the probability distribution of $x$ in state $n$ at time $t$, $P_n(x,t)$, reads
\begin{align}
	\frac{\partial P_n}{\partial t}
	&= \frac{1}{\gamma}\frac{\partial}{\partial x}
	\left(k_\mathrm{B}T\frac{\partial}{\partial x} + \frac{dV_n}{dx} - F\right)P_n
	\nonumber\\
	&\quad{}-(w_n^{+} + w_n^{-})P_n 
	+ w_{n-1}^{+}P_{n-1} + w_{n+1}^{-}P_{n+1}\,.  
\label{eq:fpeq}
\end{align}

In this work, we restrict ourselves to a particular case of 
\begin{equation}
\label{eq:vw}
	V_0(x) = \frac{1}{2}Kx^2\,, 
	\qquad
	w_0^{+}(x) = k\exp(ax)\,, 
\end{equation}
for simplicity, where $K$, $k$, and $a$ are positive constants. 
One of the reasons why the exponential dependence of 
$w_0^{+}$ on $x$ in Eq.~(\ref{eq:vw}) has been chosen is that 
a similar dependence was observed for F$_1$-ATPase~\cite{watanabe12, adachi12}. 
From Eqs.~(\ref{eq:dbalance}) and (\ref{eq:vw}), we find that 
$w_0^{-}(x)$ also depends exponentially on $x$. 
It should be remarked that a model equivalent to our model with $V_0$ and $w_0^{+}$ 
given by Eq.~(\ref{eq:vw}) was studied in Ref.~\citen{kawaguchi14} for 
a different purpose. 

A comment concerning the modeling of the linear motor may be in order. 
In motor proteins, such as kinesin and myosin V, which walk on linear tracks, 
the forward step is associated with ATP hydrolysis, 
but the backward step is usually not caused by its reverse reaction 
(ATP synthesis)~\cite{nishiyama02, carter05, gebhardt06}. 
Hence, our model as it is may not be applicable to these motor proteins. 
However, since the diffusion enhancement to be discussed in this work can occur
in the situation where the motor takes only forward steps, 
we consider that the model can be appropriate for linear motors under restricted conditions.   

%%%
\section{Velocity and Diffusion Coefficient}
\label{sec:vd}

The average velocity $v$ and the diffusion coefficient $D$ of 
the motor are defined by
\begin{equation}
\label{eq:vddef}
	v \equiv \lim_{t \to \infty}\frac{\langle x(t) - x(0)\rangle}{t},
	\quad
	D \equiv \lim_{t \to \infty}\frac{\langle [x(t) - x(0) - vt]^2\rangle}{2t}\,,
\end{equation}
where $x(t)$ is the displacement of the motor at 
time $t$ and the angular brackets indicate the statistical average. 

The average velocity can be obtained from the solution to the Fokker--Planck 
equation~(\ref{eq:fpeq}) for the steady state ($\partial P_n/\partial t = 0$) 
as follows. 
Note that the steady-state distribution $P_n(x)$ has 
the same periodicity, $P_n(x) = P_0(x - n l)$, 
as $V_n(x)$ and $w_n^\pm(x)$. 
Let $P(x)$ be the rescaled $P_0(x)$ such that $\int_{-\infty}^\infty P(x)\,dx =1$. 
Then we have 
\begin{equation}
\label{eq:v}
	v = \frac{1}{\gamma}\int_{-\infty}^\infty f(x)P(x)\,dx\,,
	\quad
	f(x) \equiv F - \frac{dV_0}{dx}.
\end{equation}

To calculate the diffusion coefficient, 
we need to solve a differential equation, which is related to the 
Fokker--Planck equation for the steady state~\cite{harms97, sasaki03}: 
\begin{align}
\label{eq:qeq}
	&\frac{1}{\gamma}\frac{d}{dx}\left[k_\mathrm{B}T\frac{d}{dx} 
	- f(x)\right]Q(x)
	- [w_0^{+}(x) + w_0^{-}(x)]Q(x)
	\nonumber\\
	&\quad{}+ w_0^{+}(x+l)Q(x+l) + w_0^{-}(x-l)Q(x-l)
	\nonumber\\
	&= \left[v - \frac{1}{\gamma}\left(f(x) - 2k_\mathrm{B}T\frac{d}{dx}\right)\right]
	P(x)\,.
\end{align}
Note that if the right-hand side of this equation were set to zero,  
the resulting equation with $Q(x)$ replaced by $P(x)$ would be identical to the 
equation for the steady-state distribution $P(x)$ obtained from Eq.~(\ref{eq:fpeq}). 
From a solution $Q(x)$ of Eq.~(\ref{eq:qeq}), the diffusion coefficient $D$ is 
obtained as 
\begin{equation}
\label{eq:dformula}
	D = D_0 + \int_{-\infty}^\infty\left[f(x)/\gamma - v\right]Q(x)\,dx\,,
\end{equation}
where
\begin{equation}
\label{eq:d0}
	D_0 = k_\mathrm{B}T/\gamma
\end{equation}
is the diffusion coefficient due to 
the Einstein relation. 
It is remarked that if $Q(x)$ is a solution to Eq.~(\ref{eq:qeq}), then 
$Q(x) + cP(x)$, with $c$ an arbitrary constant, is also a solution. 
However, this ambiguity in $Q(x)$ does not affect the result for $D$. 

%%%
\section{Diffusion Enhancement}
\label{sec:diffusion}

It is convenient to work with dimensionless variables and parameters 
to present the results succinctly. 
By introducing the dimensionless time $\tau$ and displacement $\xi$ 
defined by 
\begin{equation}
\label{eq:tx}
	\tau \equiv \frac{D_0}{l^2}t\,, \quad
	\xi \equiv \frac{x}{l} + \frac{\Delta\mu}{Kl^2}\,,
\end{equation}
and the dimensionless external force $\tilde F$ and potential $U_n$ defined by 
\begin{equation}
\label{eq:uk}
	\tilde F \equiv \frac{Fl + \Delta\mu}{k_\mathrm{B}T}\,,
	\quad
	U_n(\xi) \equiv \frac{\tilde K}{2}(\xi - n)^2\,,
	\quad
	\tilde K \equiv \frac{Kl^2}{k_\mathrm{B}T}\,, 
\end{equation}
the Fokker--Planck equation~(\ref{eq:fpeq}) with Eq.~(\ref{eq:vw}) is rewritten as
\begin{align}
	\frac{\partial P_n}{\partial \tau}
	&= \frac{\partial}{\partial \xi}
	\left(\frac{\partial}{\partial \xi} + \frac{dU_n}{d\xi} - \tilde F\right)P_n
	\nonumber\\
	&\quad{}-(\omega_n^{+} + \omega_n^{-})P_n 
	+ \omega_{n-1}^{+}P_{n-1} + \omega_{n+1}^{-}P_{n+1}\,,  
\label{eq:fpeq1}
\end{align}
where the dimensionless transition rates $\omega_n^{\pm}$ are given by
\begin{equation}
\label{eq:wpm}
	\omega_n^\pm(\xi) \equiv \kappa^\pm\exp[\pm\alpha^\pm(\xi - n)]
\end{equation}
with 
\begin{equation}
\label{eq:kap}
	\kappa^{+} \equiv \frac{kl^2}{D_0}
	\exp\left(-\frac{a\Delta\mu}{Kl}\right)\,,
	\quad
	\alpha^{+} \equiv al
\end{equation}
and
\begin{equation}
\label{eq:kam}
	\kappa^{-} \equiv \kappa^{+}\exp(\alpha^{+} - \tilde K/2)\,,  
	\quad
	\alpha^{-} \equiv \tilde K - \alpha^{+}\,.
\end{equation} 
It is not difficult to see from Eqs.~(\ref{eq:v})--(\ref{eq:kam}) 
that the dimensionless velocity $\tilde v$ and diffusion coefficient $\tilde D$ 
defined by 
\begin{equation}
\label{eq:vdtilde}
	\tilde v = vl/D_0\,,
	\quad 
	\tilde D = D/D_0
\end{equation}
depend on 
only four parameters, $\tilde K$, 
$\alpha^{+}$, $\kappa^{+}$, and $\tilde F$. 

It is remarked that $\tilde v$ and $\tilde D$ do not depend explicitly on 
the free energy change $\Delta \mu$ associated with the chemical reaction. 
This is because a particular set of the potential $V_0(x)$ and transition rate $w_0^{+}(x)$ 
given in Eq.~(\ref{eq:vw}) has been chosen. 
In this situation, $\Delta\mu$ can be absorbed into the dimensionless displacement $\xi$, 
the effective external force $\tilde F$, and the effective rate constant $\kappa^{+}$ as 
given in Eqs.~(\ref{eq:tx}), (\ref{eq:uk}), and (\ref{eq:kap}), respectively. 
This helps us to reduce the number of parameters to work with.

The Fokker--Planck equation~(\ref{eq:fpeq1}) for the steady state and the 
dimensionless version of Eq.~(\ref{eq:qeq}) were solved numerically 
by the finite-difference method with a grid spacing of $\Delta\xi = 0.005$ or smaller. 

\begin{figure}[htb]
\centerline{\includegraphics[width = 7cm]{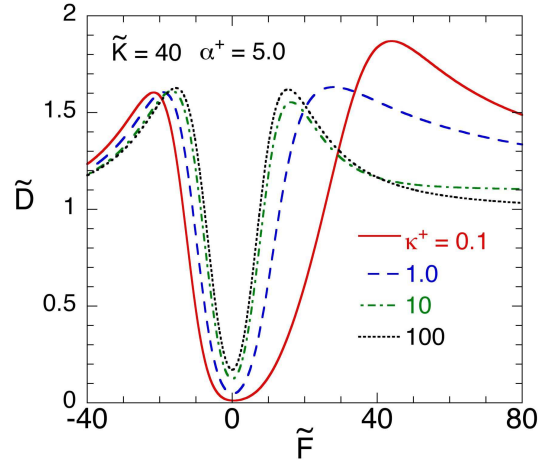}}
\caption{(Color online) 
	Dependence of the diffusion coefficient 
	$\tilde D$ on the external force $\tilde F$ 
	for $\tilde K = 40$ and $\alpha^{+} = 5$ for different 
	choices of rate constant $\kappa^{+}$ as indicated in the figure. 
	The dimensionless quantities $\tilde D$, $\tilde F$, $\tilde K$, $\alpha^{+}$, and 
	$\kappa^{+}$ are defined in Eqs.~(\ref{eq:uk}), (\ref{eq:kap}), and (\ref{eq:vdtilde}). 
}
\label{fig:df}
\end{figure}

Our primary interest is in the dependence of $\tilde D$ on $\tilde F$ and 
how this dependence changes with $\tilde K$, $\alpha^{+}$, 
and $\kappa^{+}$. 
Figure~\ref{fig:df} shows examples of such dependence for 
$\tilde K = 40$ and $\alpha^{+} = 5$ 
with several choices of $\kappa^{+}$. 
In each example, 
we observe two peaks, one for positive $\tilde F$ and 
one for negative $\tilde F$. 
For the external force $\tilde F$ around these peaks, 
the diffusion is \textit{enhanced} ($\tilde D > 1$) 
compared with the diffusion expected from the Einstein relation ($\tilde D = 1$). 
It is also noted that $\tilde D(\tilde F)$ is almost symmetric about $\tilde F = 0$  
for $\kappa^{+} = 100$ 
and is asymmetric for other cases; 
the asymmetry is more prominent for smaller $\kappa^{+}$. 
This asymmetry arises from the asymmetry between the forward and backward 
transition rates $\omega_n^{\pm}(\xi)$;  
if $\alpha^{+} = \tilde K/2$, 
then we have the symmetry relation $\omega_n^{-}(\xi) = \omega_{-n}^{+}(-\xi)$ 
and it is not difficult to see that $\tilde D(\tilde F) = \tilde D(-\tilde F)$ holds 
in this particular case. 
We expect that $\tilde D(\tilde F) \not= \tilde D(-\tilde F)$ unless 
$\alpha^{+} = \tilde K/2$. 
The reason why $\tilde D(\tilde F)$ is symmetric for large $\kappa^{+}$ 
even if $\alpha^{+} \not= \tilde K/2$ will be  
discussed in Sect.~\ref{sec:largek}. 

\begin{figure}[t]
\centerline{
	\includegraphics[scale = 0.6]{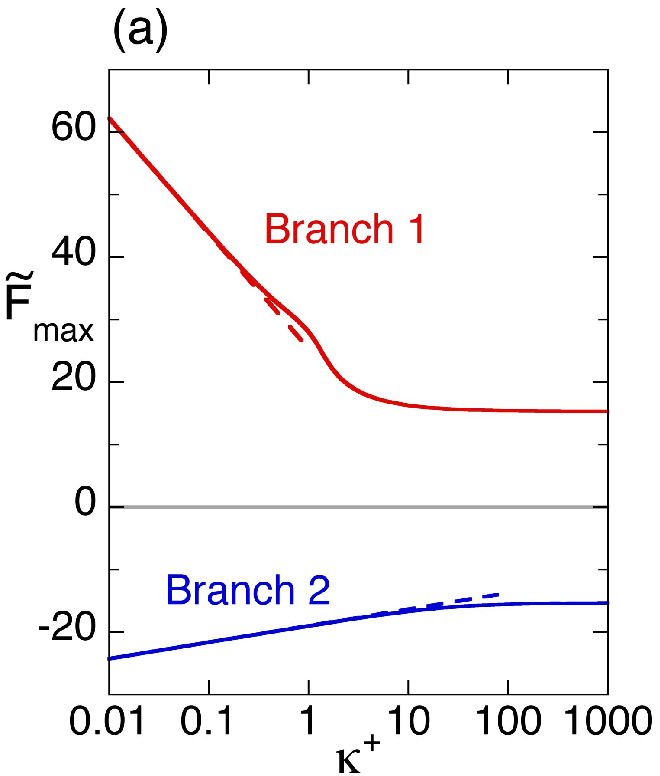}
	\hspace{5pt}
	\includegraphics[scale = 0.6]{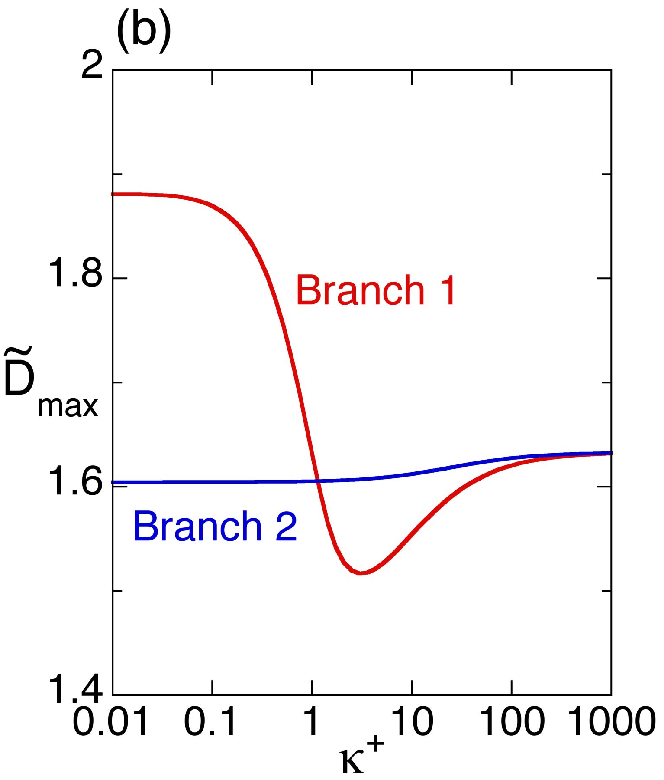}
}
\caption{(Color online) 
	(a) Position $\tilde F_\mathrm{max}$ and 
	(b) height $\tilde D_\mathrm{max}$ of the peaks in the graphs 
	shown in Fig.~\ref{fig:df} plotted against $\kappa^{+}$. 
	The dashed lines indicate the asymptotic behaviors of $\tilde F_\mathrm{max}$ 
	expressed as Eq.~(\ref{eq:fmax}). 
}
\label{fig:fmax}
\end{figure}

Figure~\ref{fig:fmax} shows 
how the position and the height of the peak observed in Fig.~\ref{fig:df} 
vary with $\kappa^{+}$: the value of $\tilde F$ at which $\tilde D$ reaches 
a maximum is denoted by $\tilde F_\mathrm{max}$, 
and the corresponding maximum value is denoted by $\tilde D_\mathrm{max}$. 
The peak for $\tilde F > 0$ in Fig.~\ref{fig:df} is referred to as \textit{branch~1}, 
and the one for $\tilde F < 0$ is referred to as \textit{branch~2}. 
In the limit of large $\kappa^{+}$, 
$\tilde F_\mathrm{max}$ and $\tilde D_\mathrm{max}$ tend to certain limiting values 
for both branches~1 and 2. 
The limiting values of $\tilde F_\mathrm{max}$ for branches~1 and 2 are 
the same in their magnitudes and opposite in their signs, 
whereas those of $\tilde D_\mathrm{max}$ are identical. 
These properties are consistent with the symmetry of $\tilde D(\tilde F)$ 
for large $\kappa^{+}$ mentioned above. 
As we will show in Sect.~\ref{sec:ka}, 
in the limit of small $\kappa^{+}$, the asymptotic behavior of $\tilde F_\mathrm{max}$ 
can be expressed as 
\begin{equation}
\label{eq:fmax}
	\tilde F_\mathrm{max} \sim \pm(\tilde K/\alpha^\pm) \left[\lambda(\tilde K, \alpha^\pm) 
	- \ln \kappa^\pm\right], 
\end{equation}
where the upper and lower signs are for branches~1 and 2, respectively, 
and $\lambda(\tilde K,\alpha)$ is a function of $\tilde K$ and $\alpha$; 
the dashed lines in Fig.~\ref{fig:fmax}(a) represent this expression 
with $\lambda \approx 3.17$ for branch 1 and $\lambda \approx 1.62$ for branch 2. 
On the other hand, $\tilde D_\mathrm{max}$ converges to definite values in this limit. 
These properties will be discussed in Sects.~\ref{sec:smallk} and \ref{sec:ka}. 
In the intermediate region of $\kappa^{+}$, 
we notice peculiar behaviors of $\tilde F_\mathrm{max}$ 
and $\tilde D_\mathrm{max}$ for branch~1: 
the graph of $\tilde F_\mathrm{max}$ has inflection points 
and that of $\tilde D_\mathrm{max}$ has a dip. 
We do not have intuitive explanations for these behaviors. 

%%%
\subsection{Limit of large $\kappa^{+}$}
\label{sec:largek}

\begin{figure}[hbt]
\centerline{
	\includegraphics[width = 7cm]{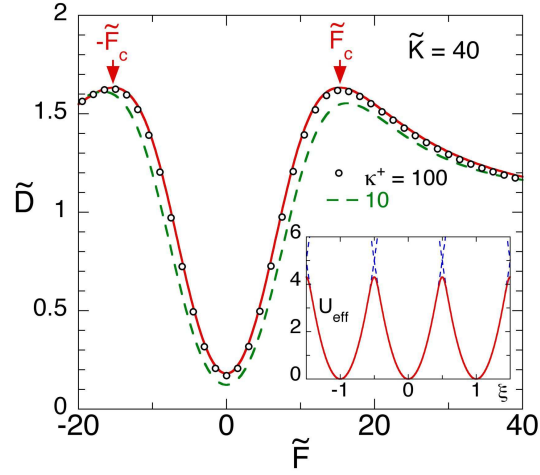}
}
\caption{(Color online) 
	Diffusion coefficient (solid line) as a function of external force $\tilde F$ 
	for a particle moving in the effective potential 
	$U_\mathrm{eff}(\xi)$, defined in Eq.~(\ref{eq:ueff}),  
	for $\tilde K = 40$ shown in the inset (solid line). 
	The data from Fig.~\ref{fig:df} for $\kappa^{+} = 100$ and 10 are also shown 
	(open circles and dashed line, respectively). 
	The vertical arrows indicate the locations of $\pm \tilde F_\mathrm{c}$, 
	where $\tilde F_\mathrm{c}$ is the maximum slope of the effective potential. 
	The dashed lines in the inset are the potentials $U_n(\xi)$ ($n = 0, \pm1,\pm2, \dots$) 
	given in Eq.~(\ref{eq:uk}). 
}
\label{fig:dfhighrates}
\end{figure}

Let us discuss the case of large $\kappa^{+}$. 
If the transition rates $\omega_n^\pm$ are large enough, 
the motor settles in a ``chemical equilibrium'' before the displacement $\xi$ 
changes appreciably. 
In this limiting case, the motor at displacement $\xi$ is in state $n$ with 
probability 
\begin{equation}
\label{eq:phin}
	\phi_n(\xi) = e^{-U_n(\xi)}
	\left[\sum_{m = -\infty}^\infty e^{-U_m(\xi)}\right]^{-1}\,,
\end{equation}
where $U_n(\xi)$ is the dimensionless potential defined in Eq.~(\ref{eq:uk}). 
Hence, the motor can be described as a Brownian particle moving in an  
\textit{effective potential}\cite{kawaguchi14}
\begin{equation}
\label{eq:ueff}
	U_\mathrm{eff}(\xi) = \sum_{n = -\infty}^\infty\int 
	\frac{dU_n(\xi)}{d\xi}\phi_n(\xi)\,d\xi\,,
\end{equation}
which is a periodic function of period 1 and is symmetric, 
$U_\mathrm{eff}(\xi) = U_\mathrm{eff}(-\xi)$. 
An example of $U_\mathrm{eff}(\xi)$ is depicted in the inset of 
Fig.~\ref{fig:dfhighrates}. 
Note that $U_\mathrm{eff}$ is almost identical to one of the $U_n$, 
if $\tilde K$ is large ($e^{\tilde K} \gg 1$), 
except for small intervals of width on the order of $1/\tilde K$ 
around $\xi = n + 1/2$ ($n = 0, \pm1, \pm2, \dots$) 
where the graphs of $U_n(\xi)$ and $U_{n+1}(\xi)$ intersect. 
This observation indicates that the transitions occur mainly when the motor 
is in these small intervals, 
where we have 
$\omega_n^{+} \approx \omega_{n+1}^{-} \approx \kappa^{+}\exp(\alpha^{+}/2)$. 
Therefore, the condition for the description in terms of the effective potential 
to be appropriate should be as follows~\cite{kawaguchi14}: 
these rates are large compared with the inverse of the time ($\sim 1/\tilde K^2$) 
for the particle to traverse an interval of length $1/\tilde K$ by diffusion 
or by drift with velocity $\tilde K/2$ (due to 
$U_n$ or $U_{n+1}$ at $\xi \sim n + 1/2$), \textit{i.e.}, 
\begin{equation}
\label{eq:cond}
	\kappa^{+} \gg \tilde K^2\exp(-\alpha^{+}/2).
\end{equation}

The diffusion of a Brownian particle moving in a one-dimensional 
periodic potential under a constant external force was studied thoroughly  
by Reimann \textit{et al.}~\cite{reimann01, reimann02}, 
and a closed-form expression for the diffusion coefficient was 
obtained~\cite{reimann01, reimann02, hayashi04, sasaki05}. 
In the present context, it is given by 
\begin{equation}
\label{eq:dreimann}
	\tilde D = \left[\int_0^1 I_{-}^2(\xi)I_{+}(\xi)\,d\xi\right]
	\left[\int_0^1I_{-}(\xi)\,d\xi\right]^{-3}\,,
\end{equation}
where 
\begin{equation}
	I_\pm(\xi) \equiv \int_0^1\exp\left[\pm U_\mathrm{eff}(\xi)
	\mp U_\mathrm{eff}(\xi \mp \eta) - \tilde F \eta\right]\,d\eta\,.
\end{equation} 
The dependence of $\tilde D$ on $\tilde F$ given by Eq.~(\ref{eq:dreimann}) 
for $\tilde K = 40$ is plotted in Fig.~\ref{fig:dfhighrates} (solid line). 
Note that $\tilde D(\tilde F)$ is symmetric about $\tilde F = 0$, 
which is due to the symmetry of $U_\mathrm{eff}(\xi)$ mentioned above. 
As explained in the introduction~\cite{reimann01, reimann02}, 
$\tilde D(\tilde F)$ has peaks near $\tilde F = \pm\tilde F_\mathrm{c}$ 
(indicated by vertical arrows in Fig.~\ref{fig:dfhighrates}), 
where $\tilde F_\mathrm{c}$ is the maximum slope 
of the effective potential $U_\mathrm{eff}$. 
In Fig.~\ref{fig:dfhighrates}, the dependence of $\tilde D$ on $\tilde F$ shown 
in Fig.~\ref{fig:df} for $\kappa^{+} = 100$ and 10 is also plotted 
(open circles and dashed line, respectively). 
We see that the data for $\kappa^{+} =100$ is very close to 
that predicted by Eq.~(\ref{eq:dreimann}), 
and that the result for $\kappa^{+} = 1000$ (data not shown) 
is indistinguishable from the latter. 
Let us check whether these numerical results are consistent with 
the condition~(\ref{eq:cond}) for the description 
in terms of the effective potential to be valid. 
The right-hand side of this inequality is about 130 
for $\tilde K = 40$ and $\alpha^{+} = 5$. 
Considering the fact that the result for $\kappa^{+} = 100$ is very close to 
the result of the effective-potential approximation, 
we suggest that the condition~(\ref{eq:cond}) can practically be replaced by 
\begin{equation}
\label{eq:wcond}
	\kappa^{+} \gtrsim \tilde K^2\exp(-\alpha^{+}/2).
\end{equation}
From all these observations, we conclude that \textit{the mechanism of  
the diffusion enhancement observed in Fig.~\ref{fig:df} 
for large $\kappa^{+}$ is essentially the same as that for the 
diffusion enhancement of the Brownian particle in a tilted periodic potential}. 

%%%
\subsection{Limit of small $\kappa^{+}$}
\label{sec:smallk}

\begin{figure}[hbt]
\centerline{
	\includegraphics[width = 6.5cm]{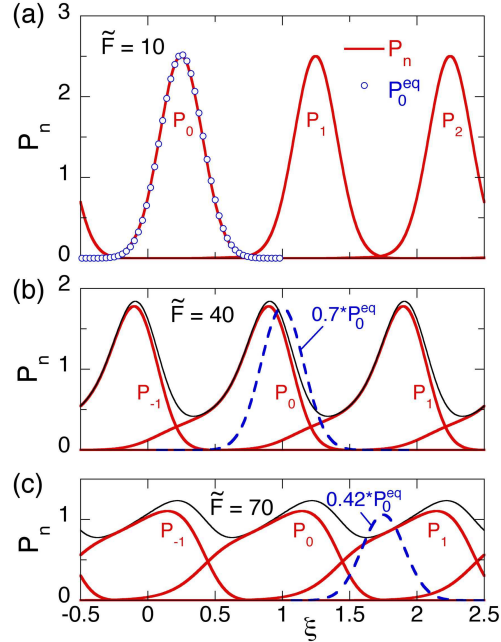}
}
\caption{(Color online) 
	Probability density functions $P_n(\xi)$ (bold solid lines) in state $n$ 
	for $\tilde K = 40$, $\alpha^{+} = 5$, and $\kappa^{+} = 0.1$ with the dimensionless force 
	(a) $\tilde F = 10$, (b) $\tilde F = 40$, and (c) $\tilde F = 70$. 
	The open circles in (a) represent the equilibrium distribution function 
	$P_0^\mathrm{eq}(\xi)$ in state $0$ in the absence of transitions to other states; 
	the dashed lines in (b) and (c) are $P_0^\mathrm{eq}(\xi)$ multiplied by $0.7$ and 
	$0.42$, respectively. 
	The thin solid lines in (b) and (c) are the total distribution functions $P_\mathrm{tot}(\xi)$ 
	defined by Eq.~(\ref{eq:ptot}). 
}
\label{fig:pn}
\end{figure}

Now we consider the case of small $\kappa^{+}$. 
If the transition rates $\omega_n^\pm$ are small enough, 
the displacement $\xi$ of the motor in state $n$ will acquire 
the equilibrium distribution  
\begin{equation}
\label{eq:pneq}
	P_n^\mathrm{eq}(\xi) \equiv \sqrt{\frac{\tilde K}{2\pi}}\,
	\exp\left[-\frac{\tilde K}{2}\left(\xi - \xi_n^\mathrm{eq}\right)^2\right]
\end{equation}
before a transition to state $n+1$ or $n-1$ takes place. 
In Eq.~(\ref{eq:pneq}), $\xi_n^\mathrm{eq}$ defined by 
\begin{equation}
\label{eq:xneq}
	\xi_n^\mathrm{eq} \equiv n + \tilde F/\tilde K
\end{equation}
is the displacement of the motor at which the external force $\tilde F$ is 
balanced with the force due to the potential $U_n(\xi)$ defined in Eq.~(\ref{eq:uk}). 
In Fig.~\ref{fig:pn}(a), 
the probability density functions (pdfs) in the steady state, 
$P_n(\xi)$ with $n = -1, 0, 1, 2$, 
are plotted together with $P_0^\mathrm{eq}(\xi)$ given by Eq.~(\ref{eq:pneq}) 
for $\tilde F = 10$ in the case of $\tilde K = 40$, $\alpha^{+} = 5$, and $\kappa^{+} = 0.1$. 
In this example, $P_0^\mathrm{eq}(\xi)$ is almost identical to $P_0(\xi)$, 
and therefore the ``quasi-equilibrium'' approximation discussed above is likely to be appropriate. 

In the quasi-equilibrium situation, the effective rates, $\omega^\pm$, 
of transitions from state $n$ are given by 
\begin{equation}
\label{eq:avomega}
	\omega^\pm
	= \!\int_{-\infty}^\infty\!\omega_0^\pm(\xi) P_0^\mathrm{eq}(\xi)\,d\xi
	= \kappa^{+}\exp\!\left[\frac{(\alpha^+)^2}{2\tilde K}
	\pm \frac{\alpha^\pm}{\tilde K}\tilde F\right]\,,
\end{equation}
and the dynamics of the motor is equivalent to 
a random walk on a lattice of lattice spacing 1 
with forward and backward hopping rates $\omega^+$ and $\omega^-$, 
respectively. 
Then, the average velocity and the diffusion coefficient 
can be expressed as 
$\tilde v = \omega^{+} - \omega^{-}$ and 
$\tilde D = (\omega^{+} + \omega^{-})/2$, respectively. 
Therefore, $\tilde D$ increases exponentially with increasing $|\tilde F|$; 
the rate of increase is determined by the factor $\alpha^\pm/\tilde K$, 
and is larger for $\tilde F < 0$ than for $\tilde F > 0$ 
in the case of $\tilde K = 40$ and $\alpha^{+} = 5$ shown 
in Fig.~\ref{fig:df}, because $\alpha^{-} = 35$ is much larger than $\alpha^{+}$ in this case. 
This prediction is consistent with the results for small $\kappa^{+}$ 
($\kappa^{+} = 0.1$ and $1.0$) presented in Fig.~\ref{fig:df}. 
Note that this exponential increase in $\tilde D$ arises 
from the fact that the transition rates $\omega_n^\pm$ increase exponentially 
with $\pm\xi$; see Eq.~(\ref{eq:wpm}). 

The condition for the quasi-equilibrium approximation for $\tilde F \sim 0$ to 
be valid may be obtained as follows. 
The relaxation time for the equilibrium in $\xi$ to be reached 
in the parabolic potential is on the order of $1/\tilde K$~\cite{risken1989}. 
If the rate $\tilde K$ of this relaxation is much larger than 
the rate of potential switching $\omega^{+} + \omega^{-}$ 
in the quasi-equilibrium approximation, 
then this approximation should be appropriate. 
Since we have 
$\omega^{+} = \omega^{-} = \kappa^{+}\exp[(\alpha^{+})^2/2\tilde K]$ for $\tilde F = 0$ 
from Eq.~(\ref{eq:avomega}), 
this condition may be expressed as
\begin{equation}
\label{eq:qecond}
	\kappa^{+} \ll (\tilde K/2)\exp\left[- (\alpha^{+})^2/2\tilde K\right].
\end{equation}
For example, the right-hand side of this inequality is about 14.6 for 
$\tilde K = 40$ and $\alpha^{+} = 5$, and 
therefore $\kappa^{+} = 1.0$ and $0.1$ satisfy this condition. 

Since $\omega^\pm$ increases exponentially with $\pm\tilde F$, 
the quasi-equilibrium approximation will no longer be valid for a large enough $|\tilde F|$: 
the forward (or backward) transition will occur  
before the equilibrium in $\xi$ is reached, 
and therefore the motor will move continuously in one direction 
without pauses around the force-equilibrium locations, 
$\xi_n^\mathrm{eq}$ given in Eq.~(\ref{eq:xneq}); 
the diffusion coefficient will be small in such a running state. 
The pdfs $P_n(\xi)$ shown in Fig.~\ref{fig:pn}(c) 
indicate that the running state seems to be realized for $\tilde F = 70$, for example,  
in the case of $\tilde K = 40$, $\alpha^{+} = 5$, and $\kappa^{+} = 0.1$:  
the function $P_0(\xi)$ is well separated from $P_0^\mathrm{eq}$ to the left, 
which suggests that the switching of potential from $U_0$ to $U_1$ occurs 
before the motor reaches the force-equilibrium position $\xi_0^\mathrm{eq}$, 
and hence the driving force $\tilde F - dU_n/d\xi$ continues to push 
the motor in one direction. 
As a result, the total pdf 
\begin{equation}
\label{eq:ptot}
	P_\mathrm{tot}(\xi) \equiv \sum_{n = -\infty}^\infty P_n(\xi), 
\end{equation}
shown by the thin solid line in Fig.~\ref{fig:pn}(c), 
does not vary much with $\xi$, 
which implies a more or less smooth flow of the motor and 
a small diffusion coefficient. 

From the above arguments, it is expected that in an intermediate range of $\tilde F$, 
the probability of the motor remaining around a force-equilibrium position and 
that of switching the potential before the motor arrives at a force-equilibrium position 
are comparable. 
The pdfs shown in Fig.~\ref{fig:pn}(b) seem to support this idea. 
In such a situation, the dispersion of the motor displacement $\xi$ 
will increase with time more rapidly than in the quasi-equilibrium state 
and in the running state. 
Therefore, the diffusion enhancement is anticipated for intermediate values of $\tilde F$. 

According to this scenario, 
the values  of $\tilde F$ around which the diffusion enhancement occurs 
may roughly be estimated as follows. 
The rate of relaxation for the equilibrium in $\xi$ to be reached 
in the parabolic potential is on the order of $\tilde K$~\cite{risken1989} as mentioned earlier. 
If the rate of potential switching is comparable to this rate of relaxation, 
the diffusion will be enhanced. 
Taking either of the effective transition rates $\omega^\pm$ in 
Eq.~(\ref{eq:avomega}) 
in the quasi-equilibrium approximation as the switching rate, 
we would have the condition $\tilde K \sim \omega^\pm$ for the diffusion enhancement. 
This means that the diffusion coefficient $\tilde D$ as a function of $\tilde F$ would 
have peaks around at $\tilde F = \tilde F^\pm$, where 
\begin{equation}
\label{eq:fmax1} 
	\tilde F^\pm \equiv \pm\frac{\tilde K}{\alpha^\pm}\left[\ln \tilde K 
	- \frac{(\alpha^\pm)^2}{2\tilde K} - \ln\kappa^\pm\right].
\end{equation}
The dependence of $\tilde F^\pm$ on $\kappa^\pm$ is similar to Eq.~(\ref{eq:fmax}) 
for the peak positions $\tilde F_\mathrm{max}$ for small $\kappa^\pm$.   
This qualitative agreement of Eq.~(\ref{eq:fmax1}) with Eq.~(\ref{eq:fmax}) 
seems to support the above scenario of the diffusion enhancement 
for small $\kappa^{+}$: \textit{the crossover from the quasi-equilibrium state to 
the running state by the force-assisted increase in the transition (switching) rate 
causes the diffusion enhancement. } 
 
%%%
\subsection{Effects of $\tilde K$ and $\alpha^{+}$}
\label{sec:ka}

So far, we have discussed how the diffusion enhancement is affected by $\kappa^{+}$. 
Now we turn our attention to the effects of $\tilde K$ and $\alpha^{+}$. 
In the case of large $\kappa^{+}$, the diffusion enhancement is essentially the 
same as in the case of a Brownian particle in a tilted periodic potential, 
as explained in Sect.~\ref{sec:largek}: 
the result does not depend on $\alpha^{+}$, 
and the peak position $\tilde F_\text{max}$ in $\tilde D(\tilde F)$ is close to 
the value of $\tilde F$ where 
the effective potential $U_\text{eff}(\xi)$ has the maximum slope. 
Since it is easy to see that the maximum slope in $U_\text{eff}$ approaches 
$\tilde K/2$ as $\tilde K \to \infty$, 
we expect that $\tilde F_\text{max} \sim \tilde K/2$ for large $\tilde K$. 
In the case of small $\kappa^{+}$, however, it is not clear how the diffusion 
enhancement is affected by $\tilde K$ and $\alpha^{+}$. 
We investigate this problem in this subsection. 

As discussed in the preceding subsection, 
under the condition that the diffusion enhancement occurs 
in the case of small $\kappa^{+}$, 
the backward transition rate $\omega^{-}$ is negligibly small 
compared with the forward transition rate $\omega^{+}$ for branch 1 (positive $\tilde F$) 
or \textit{vice versa} for branch 2 (negative $\tilde F$). 
Therefore, we employ an approximation in which 
either $\omega^{-}$ or $\omega^{+}$ is neglected. 
Then, the Fokker--Planck equation~(\ref{eq:fpeq1}) is simplified as 
\begin{equation}
\label{eq:fpeq2}
	\frac{\partial \tilde P_n}{\partial\tau} 
	= \frac{\partial}{\partial \tilde\xi}\left[\frac{\partial}{\partial\tilde\xi} 
	+ \tilde K(\tilde\xi - n)\right]\tilde P_n
	- \tilde\omega_n(\tilde\xi)\tilde P_n + \tilde\omega_{n-1}(\tilde\xi)\tilde P_{n-1}\,,
\end{equation}
where $\tilde P_n \equiv P_{\pm n}$ and 
$\tilde\xi \equiv \pm(\xi - \tilde F/\tilde K)$ 
with the plus and minus signs being for branch~1 and branch~2, respectively. 
The transition rate $\tilde\omega_n(\tilde\xi)$ in Eq.~(\ref{eq:fpeq2}) 
depends on two parameters, 
\begin{equation}
\label{eq:phidef}
	\phi \equiv \pm(\alpha^\pm/\tilde K)\tilde F + \ln\kappa^\pm\,, 
	\quad
	\alpha \equiv \alpha^\pm \,,
\end{equation}
as
\begin{equation}
\label{eq:tildeomega}
	\tilde\omega_n(\tilde\xi) = \exp[\phi + \alpha(\tilde\xi - n)]\,.
\end{equation}
Note that in this approximation, the model is characterized by three parameters, $\tilde K$, 
$\phi$, and $\alpha$, 
whereas the original model is characterized by four parameters, 
$\tilde K$, $\alpha^{+}$, $\kappa^{+}$, and $\tilde F$, 
as remarked in the first paragraph of Sect.~\ref{sec:diffusion}. 
From this observation, we conclude that the dimensionless velocity $\tilde v$ and 
diffusion coefficient $\tilde D$ depend on only these three parameters: 
the external force $\tilde F$ is now absorbed into $\phi$.

\begin{figure}[htb]
\centerline{\includegraphics[width = 6.0cm]{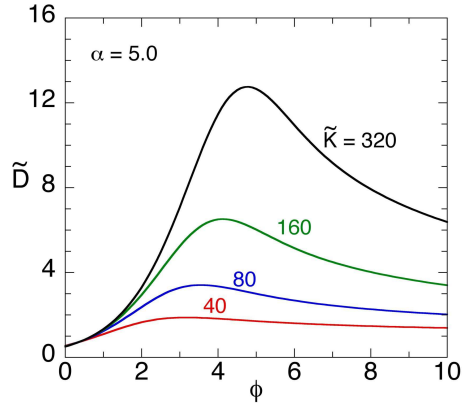}}
\caption{(Color online) 
	Dependence of the diffusion coefficient 
	$\tilde D$ on the scaled external force $\phi$ defined by Eq.~(\ref{eq:phidef})  
	for $\alpha = 5.0$ with different 
	choices of $\tilde K$ as indicated in the figure. 
}
\label{fig:dphi}
\end{figure}

Figure~\ref{fig:dphi} shows the dependence of the diffusion coefficient on 
the scaled force $\phi$ for $\alpha = 5.0$ with several choices of $\tilde K$.  
It appears that the peak height of diffusion enhancement increases with $\tilde K$. 
Let $\lambda(\tilde K, \alpha)$ be the value of $\phi$ at which 
$\tilde D(\phi)$ reaches the maximum for given $\tilde K$ and $\alpha$, 
and $\tilde D_\text{max}(\tilde K, \alpha)$ be the corresponding maximum value of $\tilde D$. 
Then, from the definition of $\phi$ in Eq.~(\ref{eq:phidef}), 
it is deduced that $\tilde F_\text{max}$ can be expressed as Eq.~(\ref{eq:fmax}) 
if either $\omega_n^{-}$ or $\omega_n^{+}$ can be neglected. 
For example, we have $\lambda \approx 3.27$ for $\tilde K = 40$ and $\alpha = 5$ 
from the result shown in Fig.~\ref{fig:dphi}, which provides the dashed line 
for branch 1 in Fig.~\ref{fig:fmax}(a). 
Note that $\tilde D_\text{max}$ depends on $\tilde K$ and $\alpha$ 
but not on $\kappa^\pm$ in the present approximation; 
this is the reason why $\tilde D_\text{max}$ in Fig.~\ref{fig:fmax}(b) tends to  
a constant value in the limit of small $\kappa^{+}$. 
The peak height of $1.88$ for $\tilde K = 40$ in Fig.~\ref{fig:dphi}, 
for example, is identical to 
the limiting value of $\tilde D_\text{max}$ for branch 1 in Fig.~\ref{fig:fmax}(b).

\begin{figure}[htb]
\centerline{
	\includegraphics[scale = 0.62]{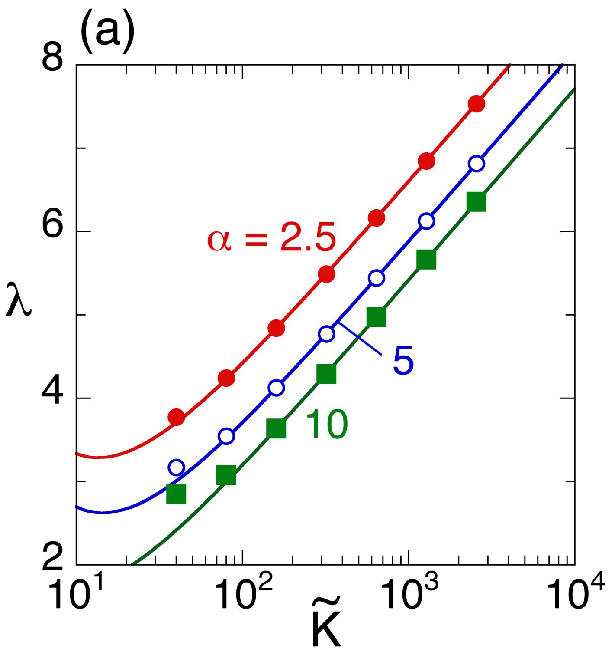}
	\hspace{2pt}
	\includegraphics[scale = 0.62]{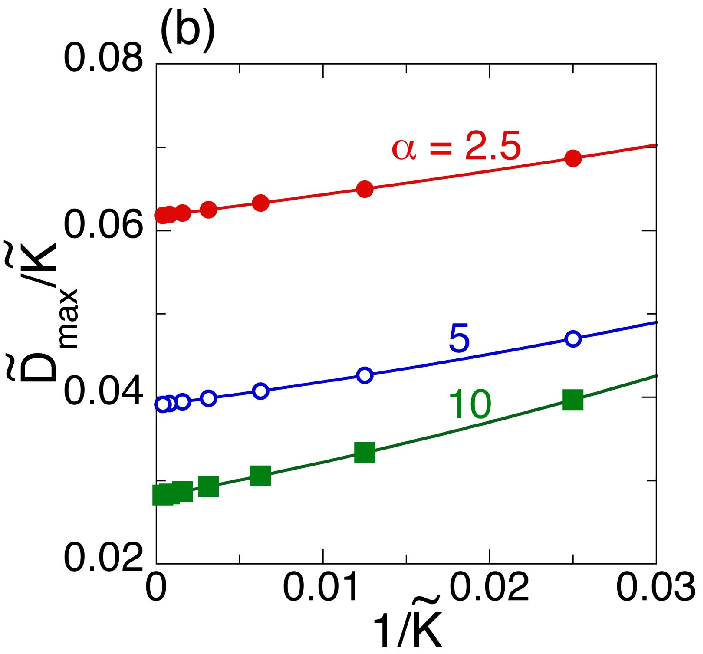}
}
\caption{(Color online) 
	Dependences of (a) $\lambda$ and 
	(b) $\tilde D_\text{max}/\tilde K$ on $\tilde K$  
	for $\alpha = 2.5$, 5, and 10. 
	Here, $\lambda(\tilde K, \alpha)$ denotes 
	the value of $\phi$ at which $\tilde D(\phi)$ reaches 
	the maximum, $\tilde D_\text{max}(\tilde K, \alpha)$, for 
	given values of $\tilde K$ and $\alpha$.  
	The lines indicate the asymptotic expressions with correction 
	terms, Eqs.~(\ref{eq:lambdak}) and (\ref{eq:dmaxk}) for $\lambda$ and 
	$\tilde D_\text{max}$, respectively. 
}
\label{fig:lambda}
\end{figure}

Now we analyze how $\lambda$ and $\tilde D_\text{max}$ depend on 
$\tilde K$ and $\alpha$. 
In Fig.~\ref{fig:lambda}(a), $\lambda$ is plotted against $\tilde K$ and 
in Fig.~\ref{fig:lambda}(b), $\tilde D_\text{max}/\tilde K$ is plotted against $1/\tilde K$ 
for $\alpha = 2.5$, 5, and 10. 
It is seen that both $\lambda$ and $\tilde D_\text{max}$ increase with $\tilde K$ 
for any value of $\alpha$, and decrease with increasing $\alpha$ for any $\tilde K$. 
The asymptotic behaviors of $\lambda$ and $\tilde D_\text{max}$ for large $\tilde K$ 
turn out to be expressed in simple mathematical forms as
\begin{equation}
\label{eq:lambda}
	\lambda \sim \ln[b(\alpha)\tilde K], 
	\quad
	\tilde D_\text{max} \sim c(\alpha)\tilde K, 
\end{equation}
where coefficients $b$ and $c$ depend on $\alpha$. 
The lines in Fig.~\ref{fig:lambda} indicate these asymptotic expressions 
with correction terms: 
we have used the expression 
\begin{equation}
\label{eq:lambdak}
	\lambda = \ln(b\tilde K) + b_1/\tilde K 
\end{equation}
for $\lambda$, where $b$ and $b_1$ are determined 
so that this expression fits   
the data of $\tilde K \ge 160$ in Fig.~\ref{fig:lambda}(a), 
and the expression 
\begin{equation}
\label{eq:dmaxk}
	\tilde D_\text{max}/\tilde K = c + c_1/\tilde K + c_2/\tilde K^2
\end{equation}
for $\tilde D_\text{max}$, where all the data in Fig.~\ref{fig:lambda}(b) are 
used to determine $c$, $c_1$, and $c_2$. 
It may be worth pointing out that the result~(\ref{eq:fmax1}) obtained from 
the qualitative argument correctly gives 
the logarithmic dependence of $\lambda$ on $\tilde K$ obtained 
numerically for large $\tilde K$.

\begin{figure}[htb]
\centerline{
	\includegraphics[scale = 0.62]{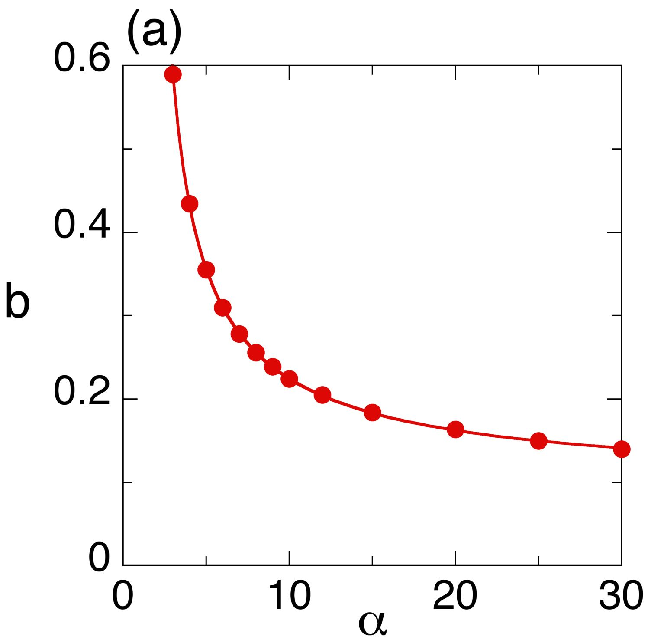}
	\hspace{2pt}
	\includegraphics[scale = 0.62]{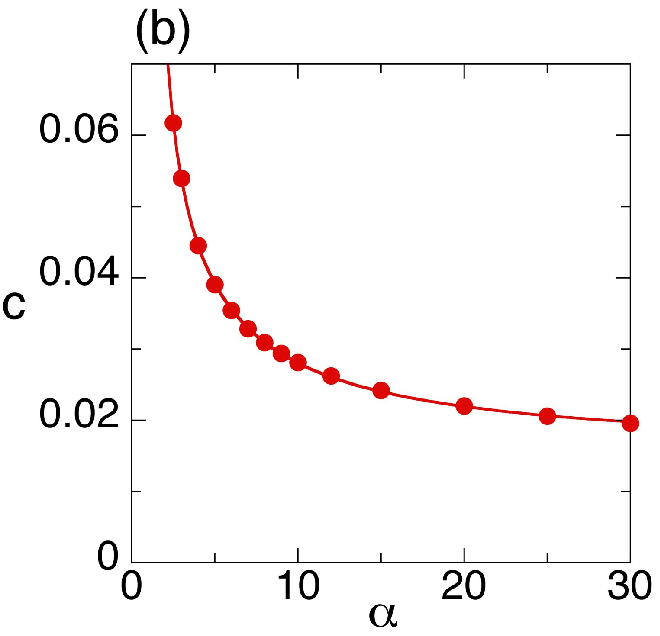}
}
\caption{(Color online) 
	Dependences of coefficients $b(\alpha)$ and $c(\alpha)$ in the asymptotic 
	expressions in Eq.~(\ref{eq:lambda}) on $\alpha$. 
	The lines represent the fitting by polynomials of degree three in $1/\alpha$. 
}
\label{fig:bc}
\end{figure}

We were not able to find simple mathematical expressions for 
the dependences of $b$ and $c$ in Eq.~(\ref{eq:lambda}) on $\alpha$. 
The numerically obtained values of $b$ and $c$ are shown as a function of $\alpha$ 
in Figs.~\ref{fig:bc}(a) and 8(b), respectively. 
These data are fitted to polynomials of degree three in $1/\alpha$, 
and the results are indicated by the solid lines in these figures. 
Both $b(\alpha)$ and $c(\alpha)$ are monotonically decreasing functions of $\alpha$ 
and appear to tend to nonzero values as $\alpha \to \infty$. 
Extrapolation using the data for large $\alpha$ yields $0.11$ and $0.015$ for 
the limiting values of $b$ and $c$, respectively. 

%%%
\section{Application to $\bf F_1$-ATPase} 
\label{sec:f1}

Our model may be too simple for describing real motor proteins. 
Nevertheless, it would be tempting to speculate whether the enhancement of diffusion 
predicted for our model can be observed in $\rm F_1$-ATPase. 
The rotational diffusion coefficient has not been measured for this molecular motor 
in the presence of both ATP and external torque, 
although the dependence of the rotation rate $\nu$ on the external torque $F$ 
was observed in the presence of ATP~\cite{watanabe-nakayama08, toyabe11}. 
Here, we determine the values of the parameters in our model 
so that the theoretical result for $\nu(F)$ agrees with the experimental data, 
and then calculate the diffusion coefficient using these parameters. 

\begin{figure}[htb]
\centerline{
	\includegraphics[width = 6.5cm]{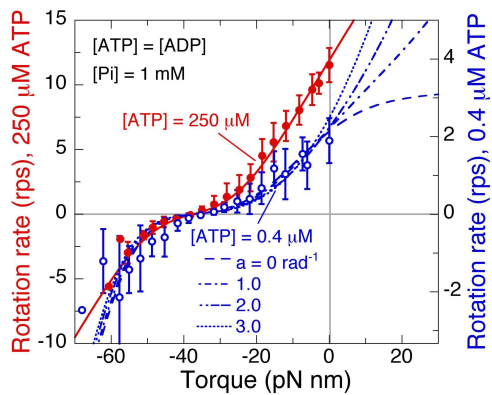}
}
\caption{(Color online) 
	Rotation rate of $\rm F_1$-ATPase as a function of 
	external torque. 
	The lines are theoretical results fitted to the experimental data (symbols) 
	obtained by Toyabe et al.~\cite{toyabe11} 
	(the error bars indicate the standard deviation). 
	Note that the scales of the vertical axes for $\rm [ATP] = 250$ and 
	$0.4\rm\,\mu M$ are different. 
}
\label{fig:f1fit}
\end{figure}

The data we use to determine the model parameters are  those shown in Fig.~\ref{fig:f1fit}, 
which were obtained by Toyabe \textit{et al}.~\cite{toyabe11}. 
Parameters other than $K$, $a$, and $k$ are determined as follows.  
We have $l = 2\pi/3$ from the threefold symmetry in the structure 
of $\rm F_1$-ATPase. 
The experimental conditions were $T = 24.5 \pm 1.5^\circ\rm C$, 
$\rm [ATP] = [ADP]$, and $\rm [Pi] = 1\,mM$, 
where [ATP], for example, stands for the ATP concentration.  
From these concentrations of substances $\Delta\mu$ may be estimated. 
However,  the estimations of $\Delta\mu$ 
are somewhat different among those published in the literature. 
We used the largest values listed in Table S1 of Ref.~\citen{toyabe11}: 
$\Delta\mu = 78.7$ and $78.4\rm\,pN\,nm$ for $\rm [ATP] = 250$ and $0.4\rm\,\mu M$, 
respectively. 
This is because, with these choices of $\Delta\mu$, good agreement is obtained between 
the stall torque (the value of torque at which the rotation rate vanishes) 
predicted by our model, $(3/2\pi)\Delta\mu$, and 
the stall torque obtained from the data in Fig.~\ref{fig:f1fit} for $\rm[ATP] = 250\rm\,\mu M$. 
The value of $D_0$ depends on the size of the probe 
(bead of $0.287\rm\,\mu m$ diameter in Ref.~\citen{toyabe11}) attached to the 
rotor ($\gamma$ subunit) and the geometry of how it is attached, 
which is difficult to control, and 
it was estimated for every measurement as described 
in Refs.~\citen{toyabe10} and \citen{toyabe12} 
to obtain 
$D_0 = 10.48$ and $11.14\rm\,rad^2/s$ for $\rm [ATP] = 250$ and $0.4\rm\,\mu M$, 
respectively~\cite{toyabe15}. 

In determining $K$, $a$, and $k$, note that 
the forward transition rate $w_n^{+}$ in the present model represents 
the following series of events in $\rm F_1$-ATPase~\cite{watanabe13a}:  
(i) the binding of ATP to the motor accompanied by the release of ADP from the motor, 
(ii) the hydrolysis of ATP catalyzed by the motor, 
and (iii) the release of Pi from the motor. 
Therefore, the rate constant $k$ in Eq.~(\ref{eq:vw}) is expected to increase 
as the ATP concentration increases. 
Here, we assume that $k$ for $\rm[ATP] = 250\,\mu M$ is 
large enough for the effective-potential description discussed in Sect.~\ref{sec:largek} 
to be appropriate; 
the validity of this assumption will be discussed at the end of this section. 
Since the rotation rate and diffusion coefficient are independent of 
$a$ and $k$ in this approximation, we can determine $K$ 
by comparing the theoretical result with the data for $\rm [ATP] = 250\rm\,\mu M$ 
in Fig.~\ref{fig:f1fit}. 
By trial and error, it was found that the data fits well to the theory with  
$K = 37\rm\,pN\,nm/rad^2$ as shown in Fig.~\ref{fig:f1fit}.
This is the same strategy as the one used by Kawaguchi \textit{et al.}~\cite{kawaguchi14} for 
their harmonic-potential model. 
However, the value $K = 47\rm\,pN\,nm/rad^2$ that they obtained is different from ours. 
The cause of this discrepancy is likely to be the different values of $D_0$ used in the 
analysis: they used $D_0 = 14.3\rm\,pN\,nm/rad^2$, 
which was reported in Ref.~\citen{toyabe12}; 
this value is different from that for the probe used to obtain the data~\cite{toyabe11} 
in Fig.~\ref{fig:f1fit}.  

We tried to determine the values of the remaining parameters $a$ and $k$ from
the data for $\rm[ATP] = 0.4\,\mu M$ in Fig.~\ref{fig:f1fit}, 
but were unsuccessful in finding a unique set of these values. 
For any value of $a$ smaller than about $3\rm\,rad^{-1}$, 
it is possible to fit the data to the theory by adjusting the value of $k$. 
Four examples of such fittings are presented in Fig.~\ref{fig:f1fit} 
for $a = 0$, 1, 2, and $3\rm\,rad^{-1}$ with 
the adjusted values of $k$ listed in Table~\ref{tab:param}. 
Although these lines lie close to each other for $F < 0$, 
they become separated as $F$ increases for $F > 0$ (assisting torque): 
the rotation rate becomes saturated for $a = 0$, 
while it keeps increasing for the other cases 
(the larger the value of $a$, the larger the rate of increase). 
In an earlier experiment~\cite{watanabe-nakayama08}, 
the rotation rate was observed to increase with assisting torque 
under various conditions, which
indicates that the value of $a$ should not be too small.

\begin{table}[t]
\caption{
The value of $k$ in the second column is determined so that the experimental data 
on the rotation rate for $\rm [ATP] = 0.4\rm\,\mu M$ shown in Fig.~\ref{fig:f1fit} 
fits the theoretical result  
for each choice of $a$ listed in the first column. 
Then, $\kappa^{+}$ and the right-hand side (RHS) of 
inequality~(\ref{eq:qecond}) are estimated for $\rm [ATP] = 0.4\rm\,\mu M$. 
Furthermore, 
$\kappa^{+}$ and RHS of inequality~(\ref{eq:cond}) are estimated 
for $\rm [ATP] = 250\rm\,\mu M$, by assuming that $k \propto \rm[ATP]$. 
}
\label{tab:param}
\begin{center}
\begin{tabular}{cc|cc|cc}
\hline
\multicolumn{1}{c}{$a$} & \multicolumn{1}{c|}{$k$} 
	& \multicolumn{2}{c|}{$\rm [ATP] = 0.4\,\mu M$}  & 
 	\multicolumn{2}{c}{$\rm [ATP] = 250\,\mu M$} \\
\cline{3-6}
\multicolumn{1}{c}{($\rm rad^{-1}$)} & \multicolumn{1}{c}{($\rm s^{-1}$)}
	 & \multicolumn{1}{|c}{$\kappa^{+}$} & \multicolumn{1}{c}{Eq.~(\ref{eq:qecond})} &
	\multicolumn{1}{|c}{$\kappa^{+}$} & \multicolumn{1}{c}{ Eq.~(\ref{eq:cond})} \\
\hline
0 & 10 & 3.94 & 19.7 & $2.62\times 10^3$ & $1.55\times 10^3$ \\
1 & 8.8 & 1.26 & 18.6 & $0.83\times 10^3$ & $0.55\times 10^3$ \\
2 & 7.2 & 0.37 & 15.8 & $0.25\times 10^3$ & $0.19\times 10^3$ \\
3 & 6.4 & 0.12  & 12.0 & $0.08\times 10^3$ & $0.07\times 10^3$ \\
\hline
\end{tabular}
\end{center}
\end{table}

Other data on the values of $a$ are available in the literature. 
First, the rate of ATP binding was measured~\cite{watanabe12, adachi12} 
as a function of the rotation angle of the rotor. 
Under low ATP concentrations, the ATP-binding event is rate-limiting~\cite{yasuda01} 
and therefore the rate of this event may be identified, 
in the first approximation,  with the rate of forward 
transition in the present model 
in analyzing the data for $\rm [ATP] = 0.4\rm\,\mu M$. 
With this identification, we obtain 
$k = 3.7\rm\,s^{-1}$ and $a = 2.6\rm\,rad^{-1}$
from the result in Ref.~\citen{watanabe12}; 
these values are not too far from those given in the last two lines in Table~\ref{tab:param}. 
Second, the theoretical investigation of Ref.~\citen{kawaguchi14} revealed that 
the \textit{asymmetry parameter} $q$, 
which corresponds to $(k_\text{B}T/Kl)a$ in the present model, 
should be close to zero so that the theory is consistent with 
the experimental finding~\cite{toyabe10} 
that the heat dissipation through the internal degrees of freedom in $\rm F_1$-ATPase 
is negligibly small. 
Note that $a = 3\rm\,rad^{-1}$ corresponds to $q \approx 0.16$, 
and therefore our choice of $a$ listed in Table~\ref{tab:param} seems consistent 
with the result of Ref.~\citen{kawaguchi14} that $q$ should be small.

From the consideration given above, we conclude that $K = 37\rm\,pN\,nm/rad^2$ 
and $0 < a < 3\rm\,rad^{-1}$ 
should be appropriate for the present model to 
describe the rotational dynamics of $\rm F_1$-ATPase. 
By using these parameters, the values of $\kappa^{+}$ defined in Eq.~(\ref{eq:kap})
and the right-hand side of inequality~(\ref{eq:qecond}) are 
estimated and listed in Table~\ref{tab:param}. 
Note that the condition for the quasi-equilibrium approximation 
to be valid [Eq.~(\ref{eq:qecond})] is satisfied for $\rm [ATP] = 0.4\rm\,\mu M$ 
for any choice of $a$ in the range given above. 

\begin{figure}[hbt]
\centerline{
	\includegraphics[width = 7.5cm]{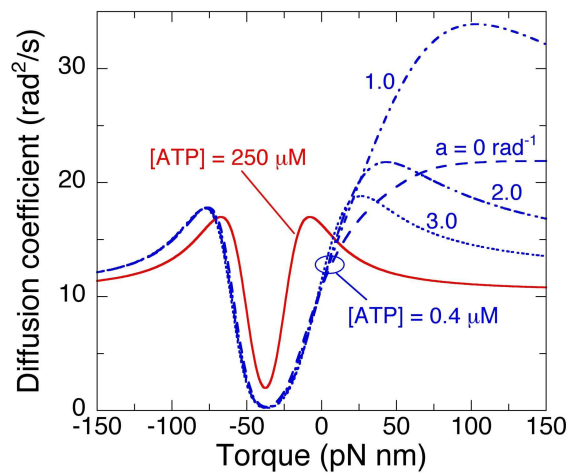}
}
\caption{(Color online) 
	Dependence of the diffusion coefficient on the external torque 
	predicted for $\rm F_1$-ATPase. 
}
\label{fig:f1df}
\end{figure}

Using the model parameters thus determined, 
we calculated the dependence of the diffusion coefficient $D$ on 
the external torque $F$, and the result is shown in Fig.~\ref{fig:f1df}.  
We see diffusion enhancements similar to the ones 
observed in Fig.~\ref{fig:df}; 
the case of $a = 0$ for $F > 0$ is exceptional, 
where $D$ increases monotonically with $F > 0$ and saturates to a limiting value. 
Since the torque of $100\rm\,pN\,nm$ is in an experimentally accessible 
range~\cite{hayashi15}, 
it is suggested that the diffusion enhancement can be observed experimentally for 
$\rm F_1$-ATPase:  
\textit{the diffusion enhancement of the type discussed in Sect.~\ref{sec:largek} 
will be observed under high ATP concentrations (e.g., $\rm [ATP] = 250\rm\,\mu M$), 
and that of the type discussed in Sect.~\ref{sec:smallk} will be observed 
under low ATP concentrations (e.g., $\rm [ATP] = 0.4\rm\,\mu M$). 
} 
However, the diffusion enhancement of the latter type under assisting torque ($F > 0$) 
may not be observed if $a$ is too small (smaller than about $1\rm\,rad^{-1}$ for 
$\rm [ATP] = 0.4\rm\,\mu M$).

Unfortunately, we were not able to determine the model parameters $a$ and $k$, 
and we could only provide ranges of these parameters, as mentioned above. 
The theoretical results shown in Fig.~\ref{fig:f1fit} indicate that 
the measurement of the rotation rate under assisting torque will help determine 
these parameters. 
In addition, the measurement of $D(F)$ will provide complementary information on $a$, 
since the position and height of the peak in $D(F)$ for $F > 0$ at low ATP concentrations 
are quite sensitive to $a$, as seen in Fig.~\ref{fig:f1df}, 
even though the precision of the diffusion coefficient determined experimentally 
may not be as good as that of the rotation rate.  

Now we discuss the validity of the effective-potential approximation used to 
analyze the data for $\rm [ATP] = 250\,\mu M$. 
The condition for this approximation to be valid is given by inequality~(\ref{eq:cond}). 
However, we do not know the value of $k$ for $\rm [ATP] = 250\,\mu M$ 
and hence cannot check whether this condition is satisfied. 
If the ATP-binding event is rate-limiting, 
the rate constant $k$ is proportional to $\rm [ATP]$ 
(since the forward transition rate can be identified with the ATP binding rate, 
which is proportional to the ATP concentration), 
and $k$ for arbitrary [ATP] can be estimated 
from the value of $k$ for $\rm [ATP] = 0.4\,\mu M$. 
Here, we tentatively assume that the relation $k \propto [\mathrm{ATP}]$ 
holds even up to $\rm[ATP] = 250\,\mu M$, 
and the values of $\kappa^{+}$ and the right-hand side of inequality~(\ref{eq:cond}) 
for $\rm[ATP] = 250\,\mu M$ estimated based on this assumption 
are listed in Table~\ref{tab:param}. 
Although these values do not satisfy the condition~(\ref{eq:cond}), 
they satisfy the weaker condition~(\ref{eq:wcond}) suggested from 
the numerical results shown in Fig.~\ref{fig:dfhighrates}. 
Therefore, the effective-potential approximation is likely to work for $\rm[ATP] = 250\,\mu M$. 
For confirmation, 
we have calculated the rotation rate and the diffusion coefficient without 
the effective-potential approximation 
for $\rm [ATP] = 250\,\mu M$ by using the parameters $K = 37\rm\,pN\,nm/rad^2$, 
$a = 0, 1, 2$, and $3\rm\,rad^{-1}$, and $k$ estimated in this way for each choice of $a$. 
The results (data not shown) are almost identical to those given in Figs.~\ref{fig:f1fit} and 
\ref{fig:f1df} obtained on the basis of the effective-potential approximation. 
Therefore, it is plausible that this approximation is appropriate. 

%%%
\section{Concluding Remarks} 
\label{sec:conclusion}

We have introduced a simple model for molecular motors to 
investigate the dependence of the diffusion coefficient on the constant external force 
in the presence of ATP, 
inspired by a similar work~\cite{hayashi15} on $\rm F_1$-ATPase in the absence of ATP. 
It turns out that the diffusion enhancement occurs, if the force is in certain ranges, 
in the presence as well as in the absence of ATP. 
The mechanism of enhancement for high ATP concentrations is essentially the same 
as that in the case without ATP (i.e., the mechanism of diffusion enhancement 
in a tilted periodic potential~\cite{reimann01, reimann02}). 
An alternative mechanism applies for low ATP concentrations, 
and the diffusion enhancement is sensitive to how the transition rate 
depends on the displacement (rotation angle) of the motor. 
It is suggested that both types of diffusion enhancement can be observed 
for $\rm F_1$-ATPase in the experimentally accessible range of external torque 
and that such observations will provide useful 
insights into the angular dependence of the reaction rate for this motor protein. 

Although the present model seems to convincingly explain the dependence of the 
rotation rate on external torque for $\rm F_1$-ATPase, 
the value of $K = 37\rm\,pN\,nm/rad^2$ we have determined is somewhat smaller than 
$K = 150\rm\,pN\,nm/rad^2$ obtained experimentally in Ref.~\citen{toyabe11} 
from the angular distribution of the probe attached to the rotor, 
or $K = 82\rm\,pN\,nm/rad^2$ estimated in Ref.~\citen{kawaguchi14} by 
analyzing the potential profile reconstructed in Ref.~\citen{toyabe12}. 
One of the reasons for this discrepancy in $K$, we suppose, 
is that the present model is too simple. 
For example, there is evidence that the interaction potential $V_n$ is not 
a simple parabola~\cite{toyabe12, kawaguchi14, gaspard07}. 
Furthermore, it may be necessary to take into account 
additional (intermediate) chemical states 
because the ATP-hydrolysis cycle catalyzed by $\rm F_1$-ATPase 
proceeds in several steps~\cite{watanabe13a}; accordingly, 
the $120^\circ$ rotation of the rotor per hydrolysis cycle 
breaks up into $80^\circ$ and $40^\circ$ substeps~\cite{yasuda01}. 
We will investigate  
a refined model, in the future, to resolve the inconsistency 
between the present theory and the experiment on $\rm F_1$-ATPase.

\begin{acknowledgments}
We thank Kumiko Hayashi and Takashi Yoshidome for useful 
discussions and comments. 
We also thank Shoichi Toyabe for providing us with the experimental data 
reported in Ref.~\citen{toyabe11} and unpublished data on $D_0$ for $\rm F_1$-ATPase. 

\end{acknowledgments}

%\appendix
%\section{}

\end{document}